\documentclass[pre,twocolumn,floatfix,showpacs]{revtex4}
\usepackage{amsmath,amsfonts,epsf,dcolumn,natbib,graphicx}
\usepackage{ascii}
\usepackage{xcolor}
\usepackage{soul}
\usepackage{ulem}
\usepackage{mathtools}
\usepackage{verbatim}
\usepackage{float}
\newcommand{\half}{\tfrac{1}{2}}

\newcommand{\be}{\begin{equation}}
\newcommand{\ee}{\end{equation}}
\newcommand{\beq}{\begin{equation}}
\newcommand{\eeq}{\end{equation}}
\newcommand{\bea}{\begin{eqnarray}}
\newcommand{\eea}{\end{eqnarray}}

\DeclareFontFamily{OT1}{pzc}{}
\DeclareFontShape{OT1}{pzc}{m}{it}%
              {<-> s * [1.180] pzcmi7t}{}
\DeclareMathAlphabet{\mathpzc}{OT1}{pzc}%
                                 {m}{it}
\begin{document}
\bibliographystyle{plainnat}
%
\title{
The importance of finite-temperature exchange-correlation\\
for warm dense matter calculations
}

\author{Valentin V.~Karasiev}
\email{vkarasev@qtp.ufl.edu}
\author{L\'azaro Calder\'{\i}n}
\author{S.B.~Trickey}
\affiliation{Quantum Theory Project, 
Dept.\ of Physics and Dept.\ of Chemistry,
University of Florida, Gainesville FL 32611-8435}

\date{13 May 2016}
\begin{abstract}
\noindent 
Effects of explicit temperature dependence in the exchange-correlation
(XC) free-energy functional upon calculated properties of matter in
the warm dense regime are investigated. The comparison is between the
KSDT finite-temperature local density approximation (TLDA) XC
functional [Phys.\ Rev.\ Lett.\ \textbf{112}, 076403 (2014)]
parametrized from restricted path integral Monte Carlo data on the
homogeneous electron gas (HEG) and the conventional Monte Carlo
parametrization ground-state LDA XC functional (Perdew-Zunger, ``PZ'')
evaluated with $T$-dependent densities.  Both Kohn-Sham (KS) and
orbital-free density functional theory (OFDFT) are used, depending
upon computational resource demands.  Compared to
the PZ functional, the KSDT functional generally lowers the
direct-current (DC) electrical conductivity of low density Al,
yielding improved agreement with experiment. The greatest lowering is
about 15\% for T= 15 kK.  Correspondingly, the KS band structure of
low-density fcc Al from KSDT exhibits a clear increase in inter-band
separation above the Fermi level compared to the PZ bands.  In some
density-temperature regimes, the Deuterium equations of state obtained
from the two XC functionals exhibit pressure differences as large as
4\% and a 6\% range of differences.  However, the Hydrogen principal 
Hugoniot is insensitive to explicit XC $T$-dependence because of
cancellation between the energy and pressure-volume work difference
terms in the Rankine-Hugoniot equation.  Finally, the temperature at
which the HEG becomes unstable is $T\geq$ 7200 K for $T$-dependent XC,
a result that the ground-state XC underestimates by about 1000 K.


\end{abstract}
\maketitle

\section{Introduction}

Warm dense matter (WDM), characterized by elevated temperatures and
wide compression ranges, plays an important role in planetary-interior
physics and materials under extreme conditions, including the path to
inertial confinement fusion, heavy ion beam experiments, and Z-pinch
compression experiments
\cite{direct-drive-ICF.2008,indirect-drive-ICF.1995,Ernstorfer.Science.2009,fast-heating-dense-plasma.2001,Hu.Militzer..2011,ZmachinePhysToday}.
Development of computational and theoretical methods to treat WDM
applications is important both for interpreting experimental results
and for gaining insight about thermodynamic regions that are difficult
to access experimentally.

\begin{figure}
\hspace{-90pt}
\includegraphics*[width=8.0cm,angle =-90]{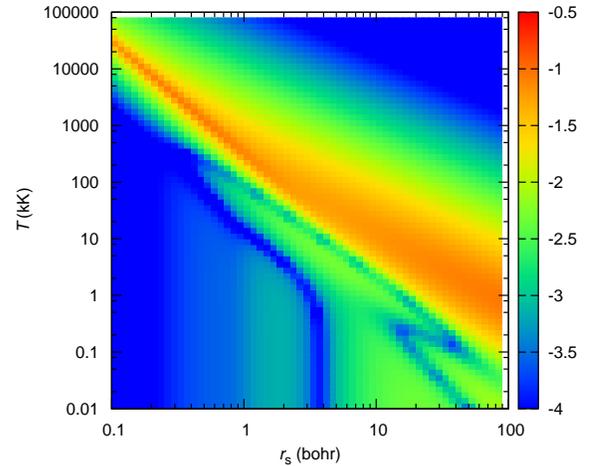}
\hspace{-90pt}
\vspace{-20pt}
\caption{
Map in $(r_{\mathrm{s}},T)$ plane which shows 
the relative importance of 
explicit $T$-dependence in the exchange-correlation
free-energy functional for the HEG
measured as 
$\log_{10}(|f_{\mathrm{xc}}(r_{\mathrm s},T)-e_{\mathrm{xc}}(r_{\mathrm s})|/[|f_{\mathrm{s}}(r_{\mathrm s},T)|+|e_{\mathrm{xc}}(r_{\mathrm s})|])$.
}
\label{fxc-map}
\end{figure}

Current practice is to treat the WDM electronic degrees of freedom via
finite-temperature density functional theory (DFT)
\cite{Mermin65,Stoitsov88,Dreizler89}. That necessitates use of an
approximate exchange-correlation (XC) free-energy density functional,
${\mathcal F}_{\mathrm xc}[n(T), T]$.  A common
approximation
\cite{Desjarlais.Kress.Collins.2002,Mazevet.Desjarlais..2005,Holst08}
is to use a ground-state XC functional evaluated with the finite-$T$
density, that is ${\mathcal F}_{\rm xc}[n({T}),{T}] %
\approx E_{\rm xc} [n({T})]$. This is the ``ground state
approximation'' or GSA.  Ref.\ \onlinecite{Burke.Entropy} 
presented a rationale for why the GSA might be expected to work well. 
The essence
of that argument is that GSA automatically fulfills certain 
constraints.  The present study gives clear demonstrations of GSA
deficiencies for specific systems in certain thermodynamic conditions 
and physical properties.  The study involves systematic investigations of three
essential questions.  What properties are affected by the explicit
$T$-dependence of ${\mathcal F}_{\mathrm xc}$, over what thermodynamic
regime does the dependence manifest itself, and what are the
magnitudes of the effects?

For compactness in what follows we use the phrase ``XC thermal
effects'' as a shortened expression for ``effects of the explicit
$T$-dependence in the XC free energy''.  Thus, XC thermal effects are
those not included in the GSA.  As orientation to the issue,
Fig.\ \ref{fxc-map} shows the relative importance of XC thermal effects
as a function of $r_{\mathrm{s}}$ (the Wigner-Seitz radius,
$r_{\mathrm s} =(3/4\pi n)^{1/3}$) and $T$ as
$\log_{10}(|f_{\mathrm{xc}}(r_{\mathrm s},T)-e_{\mathrm{xc}}(r_{\mathrm s}) %
|/[|f_{\mathrm{s}}(r_{\mathrm s},T)|+|e_{\mathrm{xc}}(r_{\mathrm s})|])$ for the homogeneous electron
gas (HEG).  $f_{\mathrm{xc}}$ is the XC free-energy per particle
\cite{LSDA-PIMC}, $e_{\mathrm{xc}}$ is the zero-$T$ XC energy per
particle \cite{PZ81}, and $f_{\mathrm{s}}$ is the non-interacting
free-energy per particle \cite{Feynman..Teller.1949}.  Note that this
ratio is the difference of energies per particle for two small
quantities divided by the energy per particle for what, in most 
cases, is a larger quantity. In particular, the denominator of the 
ratio always is greater than or equal to the magnitude of the free energy 
per particle (calculated with the zero-$T$ exchange-correlation): 
$|f_{\mathrm{s}}|+|e_{\mathrm{xc}}| \ge |f_{\mathrm{s}}+e_{\mathrm{xc}}|$.  
The ratio therefore generally underestimates the significance of  
XC thermal effects.

The orange and
yellow regions of Fig.\ \ref{fxc-map} indicate the $(r_{\mathrm s},T)$
domain wherein one may expect the $T$-dependence of XC to be important
for accurate predictions.  The nearly diagonal orange-yellow band is
particularly useful for insight.  First, it shows that finite-$T$ XC
may be expected to be important at low $T$ for large $r_{\mathrm{s}}$
values.  Second, that $T$-dependence in XC
dwindles in importance in the large-$T$ limit.  Thus the relative 
importance ratio has a
maximum at some intermediate temperature which depends on
$r_{\mathrm{s}}$. In terms of the reduced temperature, $t=T/T_{\mathrm
  F}$ ( $T_{\mathrm F}=(1/2)(9\pi/4)^{2/3}r_{\mathrm s}^{-2}$ the
Fermi temperature), the near diagonal orange-yellow band 
in Fig.\ \ref{fxc-map} is rendered in the
$(r_{\mathrm{s}},t)$ plane as a roughly horizontal band with a lower 
border starting at $t\approx 0.3$ for $r_s = 0.1$ and rising to 
$t \approx 1$ for $r_s = 100$.  That band is narrowest at low
$r_s$ and broadens by over a factor of 100 at $r_s =100$.

This plain analysis of a fundamentally important many-fermion system 
motivates investigation of XC thermal effects
upon the calculated properties of real
inhomogeneous systems.  There have been a few previous studies,
\cite{Surh2001,Renaudin..Noiret.2002,Sterne..Isaacs.2007,Faussurier..Blancard.2009,Sjostrom.Daligault.2014,DanelKazandjianPiron16}
but all except one used  $T$-dependent XC functionals
\cite{Tanaka85,Ichimaru87,PDW84,PDW2000} constructed from various approximations
to the underlying many-fermion theory, not from parametrization to 
path-integral Monte Carlo (PIMC) data.
They also involved other approximations, e.g.,
ensemble averaging of core-hole pseudopotentials in Ref.\ \onlinecite{Surh2001},
average-atom and related schemes \cite{Renaudin..Noiret.2002,Sterne..Isaacs.2007}, and 
Car-Parrinello MD in Ref.\ \onlinecite{Faussurier..Blancard.2009}.  For equations of state (EOS),
Hugoniot shock compression curves, and conductivities,
Refs.\ \onlinecite{Surh2001,Renaudin..Noiret.2002,Sterne..Isaacs.2007}
predicted significant XC temperature effects, while
Ref.\ \onlinecite{Faussurier..Blancard.2009} found only small
differences for the electrical resistivity of Aluminum.  Danel {\it et al.} 
\cite{DanelKazandjianPiron16} find consistent lowering of pressures 
from thermal XC effects and small effects on the Deuterium Hugoniot.   
The common limitation of all those studies was the 
uncontrolled nature of the local-density approximation (LDA) 
XC functionals they used.  Ref.\ \onlinecite{Sjostrom.Daligault.2014} did use
the modern Karasiev-Sjostrom-Dufty-Trickey (KSDT) finite-temperature local density approximation (TLDA) 
\cite{LSDA-PIMC}  for 
${\mathcal F}_{\mathrm xc}$ but showed
results only for the equation of state of Deuterium at relatively high material density (small-$r_{\mathrm{s}}$) 
and concluded that the fractional pressure shifts relative to ground-state
LDA were small, though not of one sign. \\ [-5pt]

In contrast, the present 
work provides an assessment of XC thermal 
effects on the basis of KSDT \cite{LSDA-PIMC} for several 
properties in diverse systems and state conditions.  
KSDT was parametrized solely to quantum Monte Carlo (QMC) plus restricted path
integral Monte Carlo (RPIMC) simulation data 
for the HEG \cite{Brown.PRL,SND.2013} and rigorous limiting 
behaviors.  KSDT therefore is the consistent counterpart to the 
widely used Perdew-Zunger (PZ) \cite{PZ81} LDA functional,
which is a parametrization of ground-state
HEG QMC data.  Lack of  consistency between the PZ parametrization 
and some earlier finite-$T$ LSDA approximations was noted explicitly as a
problem in Ref.\ \onlinecite{Faussurier..Blancard.2009}.  \\[-5pt]

The next Section gives details about the KSDT finite-$T$ and PZ XC functionals  
along with the basics of the methodology employed, including monitoring of  
entropy positivity.  Section \ref{Alconduct}
presents Kubo-Greenwood conductivity calculations on Aluminum for those
two functionals, as well as the KS band structures of fcc Aluminum 
at comparable densities and temperatures. 
Sections \ref{EOS}-\ref{hugoniot} provide 
the corresponding KSDT vs.\ PZ comparison 
for the Deuterium equation of state and for the liquid Hydrogen 
Hugoniot.
Section \ref{electron_gas} gives a brief  study of the equilibrium 
properties of the electron
gas (both HEG and with a point charge compensating background) 
at finite-$T$.  Concluding discussion is in Sec. \ref{conclusions}.

\section{Methods}
\label{method}

\subsection{Exchange-correlation free-energy functional}
\label{xc-free-energy}

To reiterate, the KSDT finite-$T$ LDA XC free-energy 
functional \cite{LSDA-PIMC} is 
a first-principles parametrization of 
RPIMC simulation data for the finite-$T$ HEG \cite{Brown.PRL}
and recent zero-$T$ QMC HEG data \cite{SND.2013}.
KSDT also has proper asymptotics 
and is free of unphysical roughness.  Additionally, it fits  
the recent data from Schoof {\it et al.}\ \cite{SchoofEtAl2015} well.  
For the spin-unpolarized XC free-energy per particle, KSDT has the form 
\be
f_{\mathrm{xc}}^{\mathrm{u}}(r_{\mathrm{s}},t)=-\frac{1}{r_{\mathrm{s}}}
\frac{a(t)+b_\mathrm{u}(t)r_{\mathrm{s}}^{1/2}+c_\mathrm{u}(t)r_{\mathrm{s}}}
{1+d_\mathrm{u}(t)r_{\mathrm{s}}^{1/2}+e_\mathrm{u}(t)r_{\mathrm{s}}}
\,.
\label{fit2}
\ee
The functions $a(t)$ and $b_\mathrm{u}(t) - e_\mathrm{u}(t)$
are tabulated in Ref.\ \onlinecite{LSDA-PIMC}.
Most calculations require evaluation of the XC free-energy, 
${\mathcal F}_{\rm xc}[n,{T}]\equiv \int d{\mathbf r} 
n({\mathbf r})f_{\mathrm{xc}}^{\mathrm{u}}(r_{\mathrm{s}}({\mathbf r}),t({\mathbf r}))$ 
and the corresponding functional derivative. 
Evaluation of properties which involve the internal energy (e.g.\ 
Hugoniot curves, heat capacities) requires the XC internal energy 
per particle as well.  It follows via the standard thermodynamic
relation ${\mathcal S}_{\rm xc}=-\partial {\mathcal F}_{\rm xc}/\partial T|_{N,V}$
as \vspace*{-6pt}
\be
\varepsilon_{\rm xc}^{\mathrm{u}}(r_{\mathrm{s}},t) = 
f_{\rm xc}^{\mathrm{u}}(r_{\mathrm{s}},t)-
t\frac{\partial f_{\rm xc}^{\mathrm{u}}(r_{\mathrm{s}},t)}{\partial t}\Big|_{r_{\mathrm{s}}}
\,,
\label{E2}
\ee
%
so the  corresponding XC internal energy is  
${\mathcal E}_{\rm xc}[n,{T}]\equiv \int d{\mathbf r} 
n({\mathbf r})\varepsilon_{\mathrm{xc}}^{\mathrm{u}}(r_{\mathrm{s}}({\mathbf r}),t({\mathbf r}))$.
Both Eqs.\ (\ref{fit2}) and (\ref{E2}) are implemented in 
our {\sc Profess@QuantumEspresso} interface 
\cite{ProfessQE,QEspresso,Ho..Carter08,Hung..Carter10,web-QTP-PP}.
(KSDT also has been implemented 
in LibXC \cite{MarquesLibXC} recently.)
Also as noted above, the comparison ground-state XC functional
evaluated with $T$-dependent densities is the well-known PZ LDA \cite{PZ81}.

\subsection{Computational details}
\label{comp-details}

Both the KSDT and PZ functionals were used in {\it ab initio}
molecular dynamics (AIMD) simulations.  We used two forms of AIMD,
with Kohn-Sham (KS) DFT forces and with orbital-free DFT (OFDFT)
forces.  For OFDFT, the non-interacting free energy functional ${\mathcal F}_s$ we used was the recently developed VT84F approximation \cite{VT84F} 
in the case of the Deuterium
equation of state and a semi-empirical ``tunable'' functional 
\cite{Tunable.2015} for Al at low material  density.  

The KS calculations used standard projector-augmented-wave (PAW)
pseudo-potential data sets \cite{atompaw-JTH-2014} (three electrons in
the valence for the Al atom), and PAWs transferable to high
compressions \cite{KarasievSjostromTrickey12B,KST2,web-QTP-PP}, all
generated with the ground state LDA XC.  For calculations with 
{\sc Profess@Q-Espresso}, that LDA XC was PZ \cite{PZ81}, while for those done
with {\sc Abinit} (see below) it was the Perdew-Wang (PW)
\cite{Perdew.Wang.1992} XC.  For the purposes of this study the
difference in behavior between those two functionals is negligible
\cite{KarasievSjostromTrickey12B}.  The PAW data sets were generated
at $T=0$ K. At the highest temperatures involved, thermal
depopulation of the core levels treated by the PAWs is minuscule.  To
illustrate, the highest LDA KS Kohn-Sham eigenvalue among the frozen
atomic Al core states is about -70 eV. At $T=30$ kK (the highest $T$
of our Al calculations), the Fermi-Dirac occupation of that level
depopulates by about $10^{-12}$. The underlying assumption (and common practice
in WDM studies) therefore is that
these PAW data sets are transferable to various thermodynamic conditions 
(i.e., the sets describe an effective core-valence interaction with 
valence electrons in various states).  Therefore, one assumes validity
for that core-valence interaction at finite-$T$ as well.
Observe that use of the $T$-dependent KSDT XC functional in subsequent 
calculations does not introduce an inconsistency because the KSDT functional 
reduces de facto to the PZ functional in the zero-$T$ limit at which the
PAW sets were generated. 

Local pseudopotentials (LPPs) 
\cite{Goodwin..Heine.1990,KarasievTrickeyCPC2012,KST2,web-QTP-PP}
developed for OFDFT and also transferable to high compressions were
used in the OFDFT calculations. For Hydrogen and Deuterium, the LPP 
only regularizes the  bare Coulomb electron-nuclear
interaction singularity, hence does not pose any possible transferability 
limitations for  high-$T$ such as those for systems with core electrons,
as just discussed.

The plane-wave energy cutoff was 500 eV for Al, and 1000 eV for Hydrogen and
Deuterium.  Further pertinent details 
are in Sec.\ \ref{EOS}.

For conductivities, we did KS-AIMD simulations 
for
$T=5$, $10$ and in some cases for $15$ kK with $\Gamma$-point-only
sampling of the Brillouin zone, the PZ XC functional, and the PAW
data set. At elevated temperatures, $T=15$, $20$, and $30$ kK, and the
low material densities of primary interest (see below), such KS-AIMD
calculations proved to be unaffordable.  In those circumstances, we
used AIMD driven by OFDFT forces from a semi-empirical ${\mathcal F}_s$ 
parametrized (``tuned'') to extrapolate KS pressure behavior
into the low material density region.  The reference for parametrization 
was KS pressure data for fcc Al at $T=8$ kK and material densities 
$0.6 \le \rho_{\rm Al} \le 2$ g/cm$^3$.  Procedural details will 
be published elsewhere \cite{Tunable.2015}.  The essential point 
here is that the AIMD
generated a sequence of ionic configurations from which a sample set
was selected (so-called ``snapshotting'') for use in standard
Kubo-Greenwood calculations \cite{Kubo.1957,Greenwood.1958}.  The
OFDFT AIMD was performed using an LDA model
LPP \cite{Goodwin..Heine.1990,KarasievTrickeyCPC2012}, again with
the {\sc Profess@Q-Espresso} interface \cite{ProfessQE,QEspresso}.
Depending on the particular material density, the AIMD was done with
16 or 32 atoms in the simulation cell such that the finite system size effects
were small 
\cite{Mazevet.Desjarlais..2005}.  
Conductivities were calculated
as averages over two to ten well-separated AIMD snapshots using a
2$\times$2$\times$2 sampling of the Brillouin zone.  The calculations used the PAW
formalism and were done with a locally modified version of {\sc
  Abinit} \cite{AbInit,ABINIT-PAW,Mazevet..Jollet.transport.2010}
which included the KSDT XC free energy functional.  We used a
3-electron PAW generated as prescribed in
Ref.\ \onlinecite{atompaw-JTH-2014}.

To gain insight and illustrate the origin of the XC-dependent differences 
in the Al DC conductivity results, a series of KS band structure calculations
was done with the same PAW data set for fcc Al with density 
$0.2$ g/cm$^3$ and $T =5$, $10$, and $20$ kK.  Those include 16,
28, and 80 bands respectively.  Those calculations were highly
converged for the fcc primitive  unit cell with
a $12\times12\times12$ Monkhorst-Pack $\textbf{k}$-grid \cite{MonkhorstPack76}.

The Hydrogen Hugoniot was studied with KS-AIMD forces 
up to $T\le 30$ kK, 
with 64 atoms in the simulation cell and a $3\times3\times3$
Monkhorst-Pack $\textbf{k}$-grid \cite{MonkhorstPack76}. Because of
computational demand issues, for $T\ge 30$ kK,
the KS-AIMD calculations used the Baldereschi mean value 
BZ point \cite{BMVP.1973}.

Finally, the various HEG stability and electron heat capacity
calculations were performed with static background (or lattice) using
KS and OFDFT respectively.

\subsection{Validation of approximate functionals for entropy positivity}
\label{entropy-constraint}

Previously, we addressed \cite{KST2} positivity of the entropy
in OFDFT for a few generalized gradient approximation (GGA) 
non-interacting free-energy functionals 
${\mathcal F}^{\mathrm{GGA}}_{\mathrm{s}}$. 
The entropy density in some cases was contaminated by local 
negative contributions. Such contamination typically leads to a 
small-magnitude contribution to the free energy compared to the total 
$T{\mathcal S}_{\mathrm{s}}$ value.  More critically,  
the global entropy value in all calculations was positive, 
consistent with the positivity constraint
being on the entropy, not on the entropy density. 

In the present work we monitored the sign of the total entropic contribution.
For the orbital-free case, that consists
of the non-interacting component 
\be
{\mathcal S}_{\mathrm{s}}[n,T]=
-\frac{\partial {\mathcal F}_{\mathrm{s}}[n,T]}{\partial T}\Big|_{N,V}
\,,
\label{S_s}
\ee
and the XC component (defined as a difference between the entropies of 
the interacting and non-interacting system 
${\mathcal S}_{\mathrm{xc}}={\mathcal S}-{\mathcal S}_{\mathrm{s}}$)

\bea
{\mathcal S}_{\mathrm{xc}}[n,T]&=&
-\frac{\partial {\mathcal F}_{\rm xc}[n,T]}{\partial T}\Big|_{N,V}
\nonumber\\
&=&\frac{1}{T}\int d{\mathbf{r}} n({\mathbf{r}})
(\epsilon^{\mathrm{u}}_{\mathrm{xc}}(r_{\mathrm{s}},t)-f^{\mathrm{u}}_{\mathrm{xc}}(r_{\mathrm{s}},t))
\,.
\label{S_xc}
\eea
In our experience, the total entropy is {\it always} positive.

\begin{figure}
\includegraphics*[width=8.0cm]{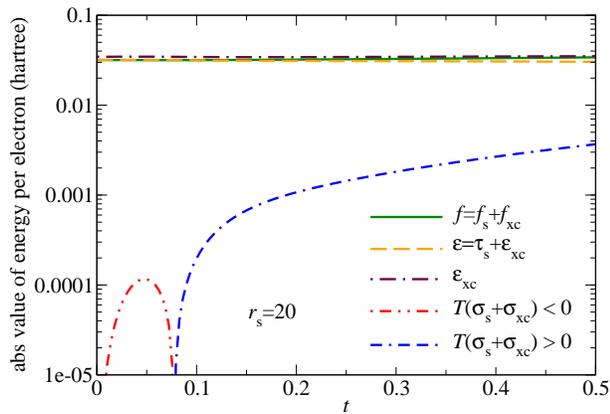}
\caption{
The total free-energy, internal energy, internal XC energy
 and entropic component (per particle) magnitudes for the
spin-unpolarized HEG for $r_{\mathrm{s}}=20$ bohr
calculated with the KSDT XC free-energy parametrization \cite{LSDA-PIMC}.
}
\label{F-HEG}
\end{figure}

For the KSDT  XC free-energy parametrization in Eq.\ (\ref{fit2}),
recently it was found \cite{Burke.Entropy} that the HEG total entropy 
becomes negative at very large
 $r_{\mathrm{s}}$ values and small temperatures
(approximately  $r_{\mathrm{s}}>10$ and $t<0.1$). Practically that 
regime is irrelevant to real systems.  
Analysis of the corresponding entropic contribution to the 
free-energy confirms that.  
Fig.\ \ref{F-HEG} shows that for $r_{\mathrm{s}}=20$ bohr
the negative entropic contribution has a maximum amplitude 
of order 0.0001 Hartree per electron. This error is negligible, since it  
is at or below the typical accuracy of finite-temperature 
Kohn-Sham and orbital-free codes. It is also negligibly small 
in comparison to the total free energy or total internal energy.
The situation is quite similar for other large-$r_{\mathrm{s}}$.
The violation is inconsequential, hence seems to be primarily of 
an aesthetic character.

After-the-fact validation of the thermal Kohn-Sham and orbital-free calculations in combination
with the KSDT XC free-energy parametrization Eq. (\ref{fit2}) show 
that the total entropy is positive for all materials and all WDM conditions 
probed in the present work.  

\section{Results}
\label{results}

\subsection{Aluminum conductivity and band structure}
\label{Alconduct}

Experimental study of the  electrical conductivity of warm dense Al 
was reported in Ref.\ \onlinecite{DeSilva.Katsouros.1998}.  Theoretical 
treatment via AIMD and the Kubo-Greenwood formula 
\cite{Kubo.1957,Greenwood.1958}
is found in Refs.\ \onlinecite{Desjarlais.Kress.Collins.2002}, 
\onlinecite{Mazevet.Desjarlais..2005}
and \onlinecite{Faussurier..Blancard.2009}.  That latter
study found the influence 
of the finite-$T$ XC functional on the DC electrical resistivity
(the inverse of electrical conductivity) to be small at material 
densities $\rho_{\rm Al}=1.0$
and $1.4$ g/cm$^3$ ($r_{\mathrm{s}}=2.89$ and 2.58 bohr respectively, assuming
the usual three free electrons) 
and $T = 5 \rightarrow 20$ kK.   The earlier studies 
\cite{Desjarlais.Kress.Collins.2002,Mazevet.Desjarlais..2005}
found that DC conductivities depend weakly 
upon $T$ in the range $6-30$ kK for  
material densities between roughly 0.5 and 2.0 g/cm$^3$.  Since
the total $T$-dependence in general is dominated by the non-interacting
free-energy contribution and the XC contribution is 
comparatively small in magnitude (recall discussion of 
Fig.\ \ref{fxc-map}), those findings mean that for this 
density range XC thermal effects should be small as well.

However, the results of 
Refs.\ \onlinecite{Desjarlais.Kress.Collins.2002,Mazevet.Desjarlais..2005}
also suggest that XC thermal effects might be noticeable 
at low material densities (between 0.025 and 0.3 g/cm$^3$).  In that
region,  the DC conductivity  has strong $T$-dependence. 
Figure \ref{fxc-map} also suggests that XC thermal effects 
should be important at such low material densities (large-$r_{\mathrm{s}}$)
for temperatures between about 10 and 500 kK. 
(At $T=15$ kK the reduced temperature is $t\approx 1.0$ and $0.6$ 
for $\rho_{\rm Al}=0.1$ and $0.2$ g/cm$^3$ respectively.)
These considerations motivated our AIMD calculations of the DC 
conductivities for three 
densities in that range, $\rho_{\rm Al}=0.1$, $0.2$, and $0.3$ g/cm$^3$ 
($r_{\mathrm{s}}=6.22$, $4.94$, and $4.21$ bohr respectively).
(Note that the foregoing $r_{\mathrm{s}}$ values are calculated with 
the conventional total number of valence electrons, 3, for Al.  
However, that could underestimate an effective free-electron
 $r_{\mathrm{s}}$ and thereby diminish the validity of correlation between 
XC thermal effects on a particular property (e.g., conductivity)
and Fig.\ \ref{fxc-map}.  Insight from that Figure depends to some
extent on how $r_{\mathrm{s}}$ for a physical system is calculated.)

The average of the Kubo-Greenwood optical conductivity over a number of 
snapshots or 
configurations ($N_{\rm config}$) as a function of frequency $\omega$ 
is given in atomic units by 
\beq
\sigma(\omega) = \frac{1}{N_{\rm config}}\sum_{I=1}^{N_{\rm config}} %
\sum_{\mathbf k}w_{\mathbf k}\sigma_{\mathbf k}(\omega; \lbrace {\mathbf R} %
\rbrace_I), 
\label{sigmaconfig}
\eeq
with
\bea
\sigma_{\mathbf k}(\omega; \lbrace {\mathbf R} \rbrace_I) &=& %
\frac{2\pi}{3\omega \Omega} \sum_{i,j}^{N_b}\sum_{\nu =1}^{3} %
(f_{i,{\mathbf k},I}- f_{j,{\mathbf k},I}) \nonumber \\
 && \hspace*{-30pt} \times \vert \langle \varphi_{j,{\mathbf k},I} %
\vert \nabla_\nu \vert \varphi_{i,{\mathbf k},I}\rangle \vert^2 %
\delta (\epsilon_{j,{\mathbf k},I}-\epsilon_{i,{\mathbf k},I} -\omega) \; .
\label{KGformulae}
\eea
Here $\Omega$ is the system volume, $w_{\mathbf k}$ is the weight of
BZ point $\mathbf k$, and  $f_{i,{\mathbf k},I}$ are Fermi-Dirac occupations of
KS orbitals $\varphi_{i,{\mathbf k},I}$.  Those orbitals and associated 
eigenvalues $\epsilon_{j,{\mathbf k},I}$ 
are indexed as band, BZ vector, and lattice configuration   
snapshot at lattice coordinates $\lbrace {\mathbf R} %
\rbrace_I$.  

The DC conductivity is the limit of $\sigma(\omega)$ as $\omega
\rightarrow 0$.  Because of the frequency-difference delta-function,
computational convergence to that limit with respect to the number of
KS bands ($N_b$) is known to be rapid
\cite{Mazevet.Desjarlais..2005}. Consequences of the numerical
implementation of the $\delta$-function are a complicating factor.
Gaussian broadening of the $\delta$-function
\cite{Desjarlais.Kress.Collins.2002} $\Delta=0.2$ eV was used.
Increasingly severe local oscillations in $\sigma(\omega)$ appear
rapidly as $\Delta$ is decreased below that value, especially at lower
temperatures.  As a consequence, the DC conductivity does not converge
as $\Delta\rightarrow 0$.  See the discussion in
Ref.\ \cite{Lambert.Recoules..Desjarlais.2011}.  The chosen value of
$\Delta$ is close to being optimal according to the criterion of
Ref.\ \cite{Desjarlais.Kress.Collins.2002} for the system size and
density-temperature range relevant here.  Admittedly, however, the
results are sensitive to that choice.  A better procedure would
determine the optimal $\Delta$ at each density and temperature.

To ensure convergence with $N_b$, our calculations used 
a minimum occupation number 
threshold of order $10^{-6}-10^{-7}$,
such that the number of bands included for $\rho_{\rm Al}=0.1$ g/cm$^3$
was $N_b=$ 208, 672, 1184, 1920, and 3096 at $T$=5, 10, 15, 20, and 30 kK respectively.
The number of bands required decreases rapidly with increasing material 
density,
but increases rapidly with increasing numbers of atoms in the 
simulation cell. The effect of these dependencies can be checked by
testing for satisfaction of the {\it f}-sum rule\cite{Kubo.1957}.  It
was  satisfied to
$90-92$\% at $T=5$ kK, and to  $95-97$\% at higher temperatures.

\begin{figure}
\includegraphics*[width=7.5cm]{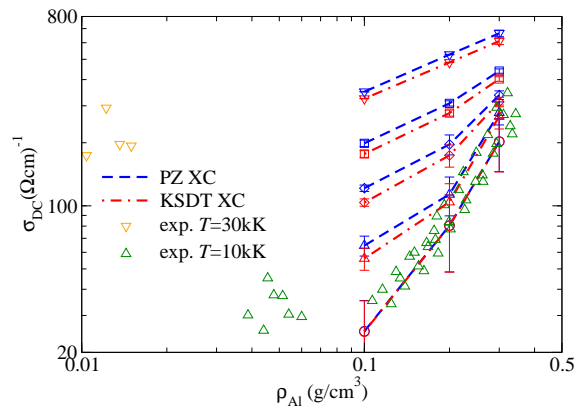}
\caption{
Aluminum DC conductivity as a function
of density from calculations with 
$T$-dependent KSDT (dot-dashed) and ground state PZ (dashed) XC functionals 
for five isotherms. From bottom to top $T=5$ kK (circles), $10$ kK (triangles up),
$15$ kK (diamonds), $20$ kK (squares) and $30$ kK (triangles down). Experimental data
\cite{DeSilva.Katsouros.1998} correspond to $10$ kK (triangles up)
and $30$ kK (triangles down).
}
\label{sigma_DC}
\end{figure}

\begin{figure}
\includegraphics*[width=7.5cm]{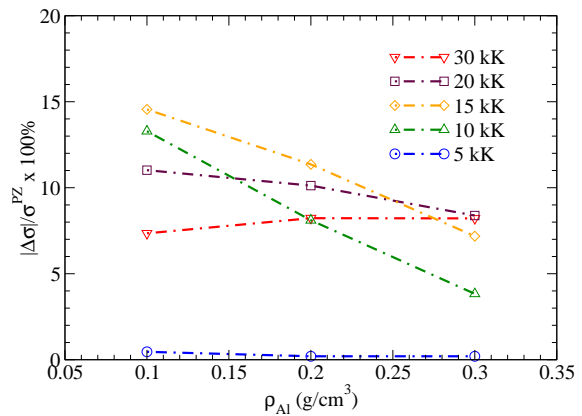}
\caption{
Relative error in DC conductivity for Al as function of density
for five different temperatures.  
}
\label{sigma_DC_relerror}
\end{figure}

Results are shown in Fig.\ \ref{sigma_DC}. The  standard deviations 
shown there as error bars correspond to averaging over the snapshots.
Note first that for all $T$, the explicitly 
$T$-dependent XC functional lowers the DC conductivity.  
Beginning at $T=5$ kK, the effect increases with increasing  
$T$ and is largest 
near $T$=15 kK, then decreases.  
Fig.\ \ref{sigma_DC_relerror} shows the relative error in using 
the ground-state XC functional
\be
\frac{\mid \Delta \sigma \mid }{\sigma^{\rm PZ}} %
:= \frac{\mid \sigma^{\rm KSDT}_{\rm DC} - \sigma^{\rm PZ}_{\rm DC}\mid}{\sigma^{\rm PZ}_{\rm DC}}  
\label{sigmaDCrelerror}  \; .
\ee
That error 
is 0.5\%, 13\%, 15\% , 11\%, and 7\% for $\rho_{\rm Al}=0.1$ g/cm$^3$ at $T=5$, $10$, $15$, $20$ and $30$ kK respectively. An important aspect is that 
the relative error is not amenable to correction by some simple, rule-of-thumb 
shift.

The number of snapshots at the lowest temperature, $T=5$ kK, is ten.  
Nevertheless 
\begin{figure}[h]
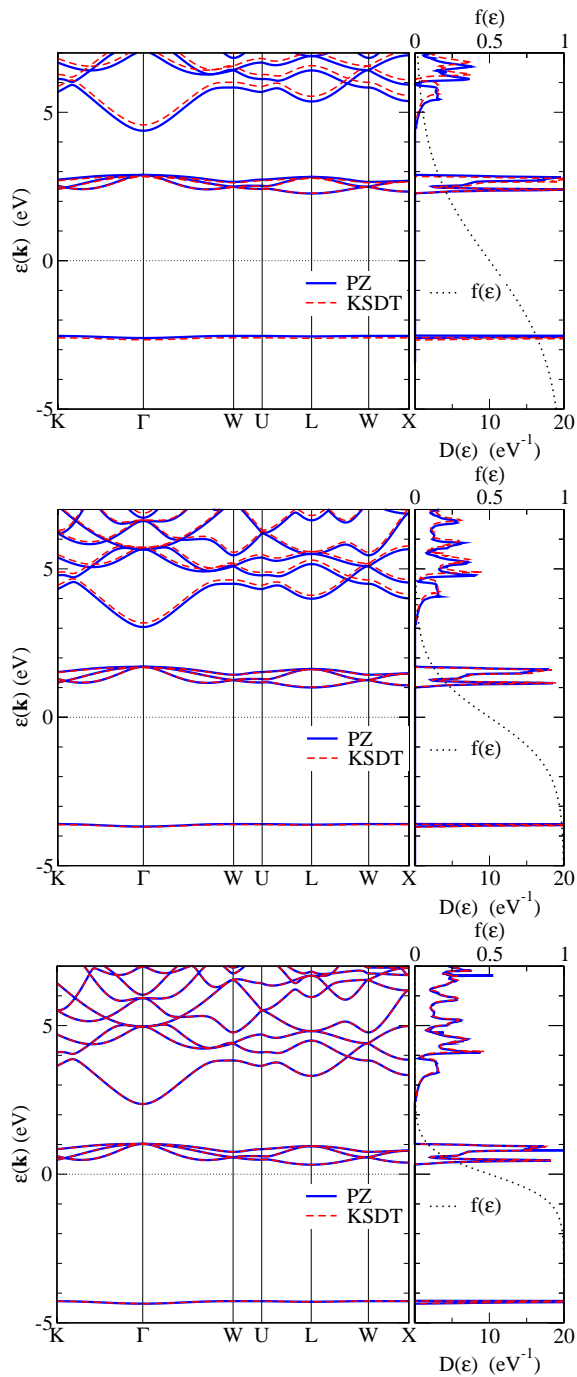

\includegraphics*[width=7.5cm]{R0.2-T20000-KP12-pz-vs-ksdt.eps}
\includegraphics*[width=7.5cm]{R0.2-T10000-KP12-pz-vs-ksdt.eps}
\includegraphics*[width=7.5cm]{R0.2-T5000-KP12-pz-vs-ksdt.eps}
\caption{
Comparison of fcc Al KS band structures for $\rho = 0.2$g/cm$^3$ from
KSDT $T$-dependent XC and PZ ground-state XC for $T = 5, 10$,
and $20$kK (bottom to top).  Fermi level $\epsilon_F$ set to 
zero.  The right-hand panels display
the density of states $D(\epsilon)$ and Fermi-Dirac occupation $f(\epsilon)$.
}
\label{KSbandsAl}
\end{figure}
the standard deviation at that $T$ is large.
To decrease it would require increasing the number of snapshots or 
the simulation cell size or both.  
Such sensitivity to the nuclear configuration may be explained by  
transient formation and dissociation of small Al clusters 
at that $T$, a process seen in the snapshots. 
Crucially, however, the difference between KSDT and PZ conductivities 
for each snapshot depends very weakly on nuclear configuration. 
Those differences, as shown in Figs.\ \ref{sigma_DC} and 
\ref{sigma_DC_relerror}, are negligible at $T=5$ kK. 
As $T$ increases, the standard deviation decreases 
as the system becomes more nearly uniform 
($10$, $8$ and $4$ snapshots were used for $T=10$, 15 and 20 kK
respectively) 
and the difference attributable to the two functionals becomes 
clearly discernible.  
Averaging over only two snapshots in fact provides very small
error bars at the highest $T=30$ kK.  

There exists also an implicit influence of $T$-dependent XC on 
DC conductivities via nuclear configurations.  That arises because a 
snapshot sequence from AIMD performed with the KSDT functional will
differ from the sequence from the PZ functional (with $T$-dependent
density of course).  To identify only the {\it explicit} dependence,
we deliberately used the same snapshots for both KSDT and PZ 
conductivity calculations.  
Evaluation of the {\it implicit} influence would require calculations of
averages over much longer snapshot sequences generated by AIMD 
with each XC functional.   
Given that it is almost certain that the implicit effects are small
compared to the explicit ones, and given the cost of doing the AIMD
calculations and snapshot conductivities (even with the cost effectiveness
of OFDFT AIMD), we opted not to pursue the implicit influence. 

Since the KS eigenvalues and orbitals are inputs to the Kubo-Greenwood
calculations, direct examination of thermal XC effects 
upon the rather unfamiliar low-density KS band-structure therefore 
is of interest.  Fig.\ \ref{KSbandsAl}
provides comparison of the fcc Al band structure at $\rho = 0.2$
g/cm$^3$ at three temperatures.  Overall there is a $T$-dependent
shifting upwards of the bands above $\epsilon_F$ as they become
increasingly occupied. For energies nearest $\epsilon_F$ on either
side, the KSDT bands lie below the PZ ones, whereas that ordering is
reversed for the bands next upward.  In those bands, at $T=20$ kK the
shift is about 0.2 eV, about 10\% of the electronic temperature.  That
is also the amount of relative increase in inter-band separation
between the band at the Fermi level and the next higher conduction
band.  The separation increase shows up as a lowering of
$\sigma(\omega)$ for small $\omega$ induced by the lowering of
Fermi-Dirac occupations and their derivatives (occupation number
difference) in Eq.\ (\ref{KGformulae}). 
$D(\epsilon)$ clearly shows not only the general shift upward that
accompanies increasing $T$, but also that the bandwidth nevertheless
is essentially unchanged.

\subsection{Equation of state of warm dense Deuterium}
\label{EOS}

To explore XC thermal effects upon bulk thermodynamics, we did KS and
OFDFT AIMD calculations on Deuterium at material densities between 0.2
and 10 g/cm$^3$ for $T = 2 \rightarrow 1000$ kK.  The familiar
unfavorable computational cost scaling with $T$ limited our KS-AIMD
results to below $T\approx$ 125 - 180 kK for higher material densities
($\rho_{\mathrm{D}}\ge 2$ g/cm$^3$), and up to $T\approx$ 60 - 90 kK
for $\rho_{\mathrm D}=0.2$ and $0.506$ g/cm$^3$.  Depending on the
material density, the simulation cells had 64, 128, or 216 atoms.
For some KS-AIMD cases, the number of atoms in the simulation cell 
was decreased with increasing $T$. 
Most of the KS calculations were performed at the $\Gamma$-point only, though
for $\rho_{\mathrm D}=0.506$ g/cm$^3$ a $2\times 2 \times 2$ 
Monkhorst-Pack Brillouin zone grid \cite{MonkhorstPack76} was used 
at the two lowest temperatures. 

The pressure converges slowly with respect to the number of MD steps, 
but pressure differences (between simulations with two different XC functionals)
typically converge more rapidly. At each density-temperature point, the
system first was equilibrated for at least 1500 MD steps, 
followed by 4500 steps 
for data gathering. The first 500 of those steps  
were considered to be an additional equilibration, hence 
were discarded during calculation of statistic averages. 
For $\rho_{\mathrm D}=0.506$ g/cm$^3$ at several temperatures, 
we also did 8500 step simulations to test pressure convergence.
The time step was scaled with increasing $T$ by a factor proportional 
to $1/\sqrt{T}$. 

\begin{figure}
\includegraphics*[width=7.5cm]{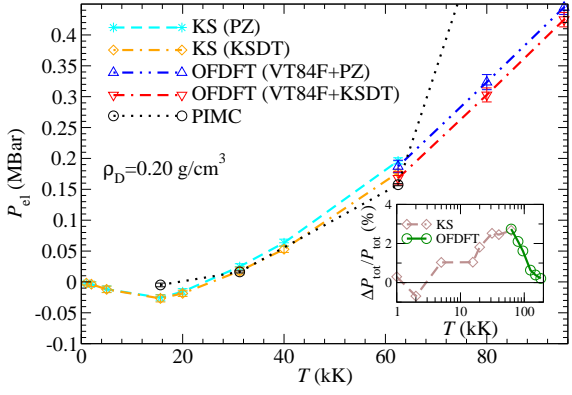}
\caption{
Deuterium electronic pressure as a function 
of $T$ from KS and OFDFT calculations 
with the finite-$T$ KSDT and ground-state PZ  
XC functionals.
Inset: relative difference between total pressure from the 
calculations with the PZ and KSDT XC; see Eq.\ (\ref{deltaP}).
System density $\rho_{\mathrm{D}}=0.20$ g/cm$^3$ ($r_{\mathrm{s}}=3$ bohr).
PIMC results are shown for comparison.
}
\label{P-T.H128R0.1}
\end{figure}

\begin{figure}
\includegraphics*[width=7.5cm]{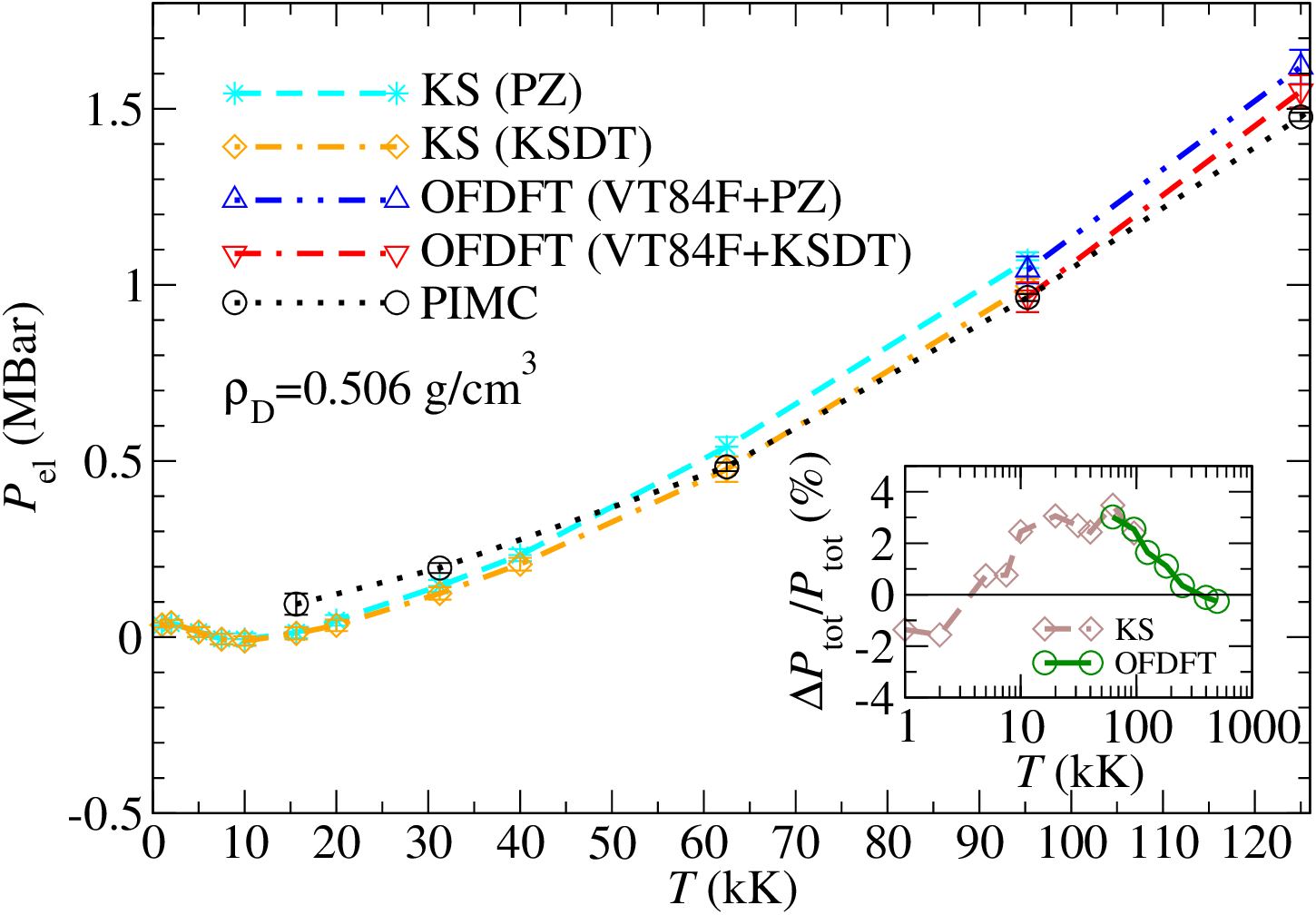}
\caption{
As in Fig.\ (\ref{P-T.H128R0.1}) for 
Deuterium, $\rho_{\mathrm{D}}=0.506$ g/cm$^3$ ($r_{\mathrm{s}}=2.2$ bohr).
}
\label{P-T.D128R0.506}
\end{figure}

\begin{figure}
\includegraphics*[width=7.5cm]{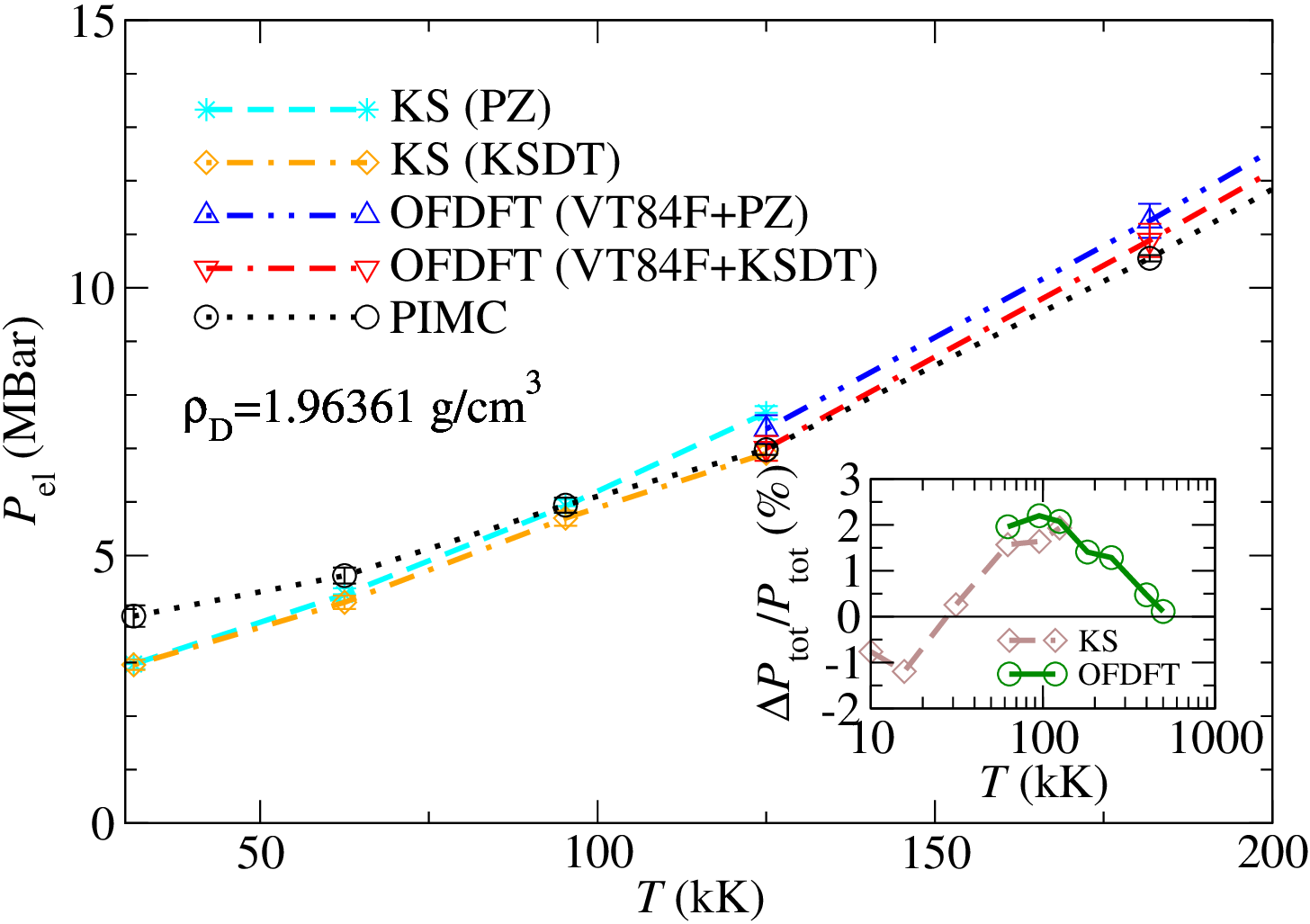}
\caption{
As in Fig.\ (\ref{P-T.H128R0.1}) for 
Deuterium, $\rho_{\mathrm{D}}=1.964$ g/cm$^3$ ($r_{\mathrm{s}}=1.4$ bohr).
}
\label{P-T.D128R1.96361}
\end{figure}

\begin{figure}
\includegraphics*[width=7.5cm]{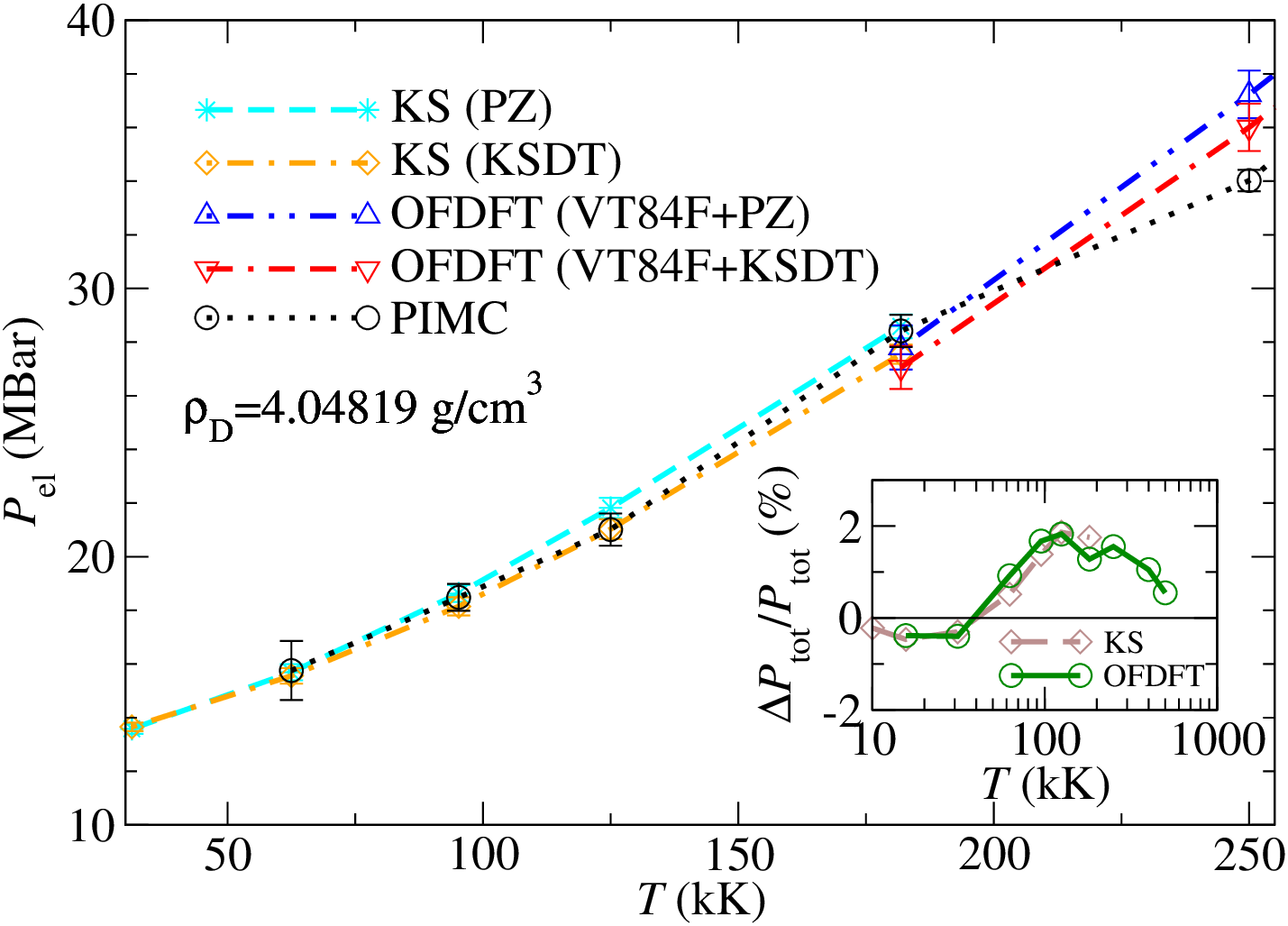}
\caption{
As in Fig.\ (\ref{P-T.H128R0.1}) for 
Deuterium, $\rho_{\mathrm{D}}=4.04819$ g/cm$^3$ ($r_{\mathrm{s}}=1.10$ bohr).
}
\label{P-T.D128R4.04819}
\end{figure}

\begin{figure}
\includegraphics*[width=7.5cm]{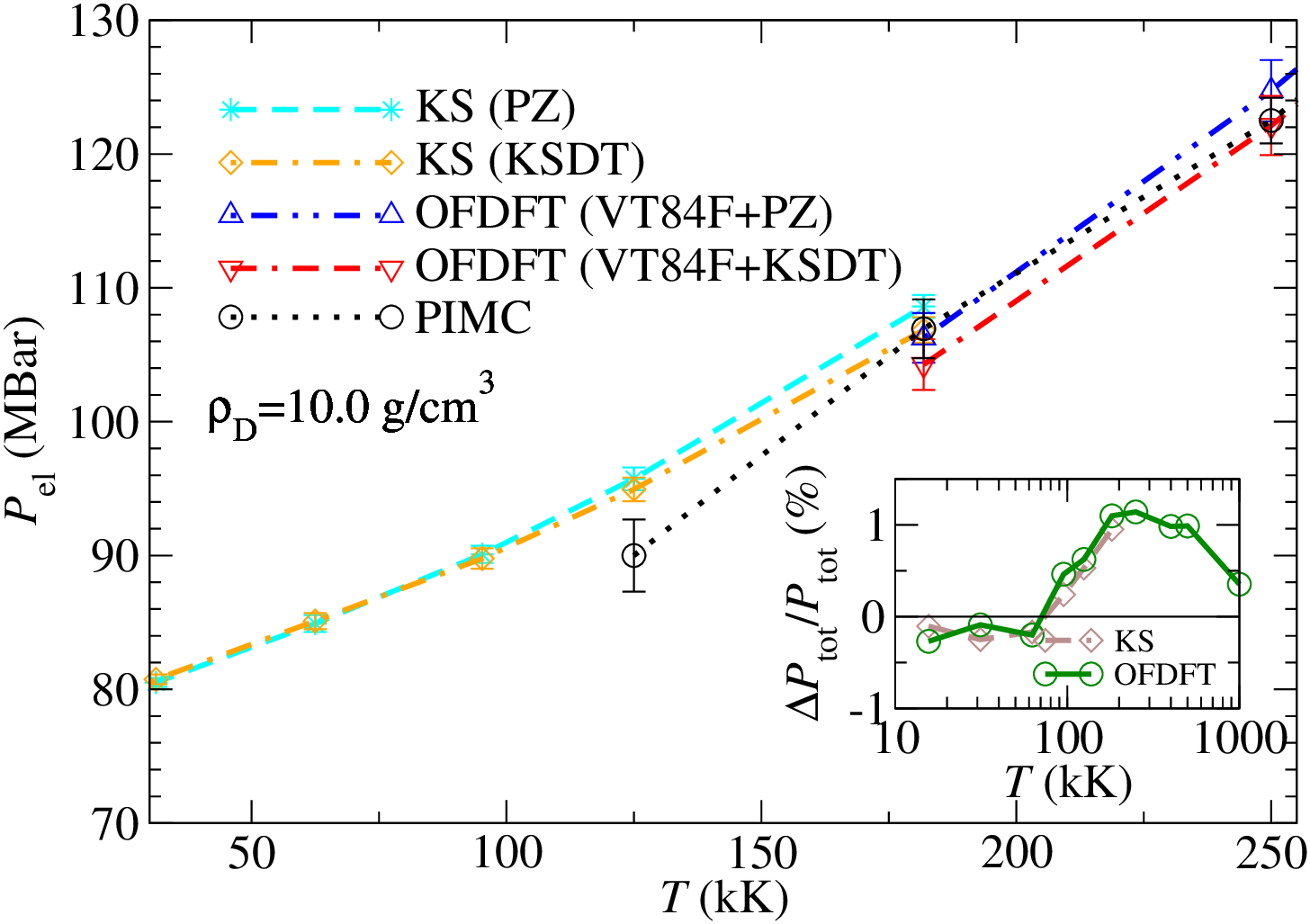}
\caption{
As in Fig.\ (\ref{P-T.H128R0.1}) for 
 Deuterium, $\rho_{\mathrm{D}}=10.0$ g/cm$^3$ ($r_{\mathrm{s}}=0.81373$ bohr).
}
\label{P-T.D216R10.0}
\end{figure}

Figures \ref{P-T.H128R0.1} - \ref{P-T.D216R10.0} compare the 
{\it electronic} pressure (that is, without the ionic ideal gas contribution)
from the KS and OFDFT calculations done in conjunction with the finite-$T$
KSDT and ground-state PZ XC functionals.  Error bars shown in those 
figures correspond to the standard deviation for the average {\it electronic} 
pressure.  The insets show the 
percentage relative difference 
for the calculated {\it total} pressures (i.e., including the thermal 
ionic contribution), namely
\be
\Delta P_{\mathrm{tot}}/P_{\mathrm{tot}}\equiv
(P_{\mathrm{tot}}^{\mathrm{PZ}}-P_{\mathrm{tot}}^{\mathrm{KSDT}})/
P_{\mathrm{tot}}^{\mathrm{PZ}}\times 100\%  \; .
\label{deltaP}
\ee
That quantity measures the XC thermal effects upon 
the {\it total} pressure in the system.

Note first that the relative difference $\Delta P_{\mathrm{tot}}$ is
of both signs, so no simple offset can be used as a correction.  
Sjostrom and Daligault \cite{Sjostrom.Daligault.2014} found pressure 
differences of both signs as well, whereas Ref.\ \cite{DanelKazandjianPiron16} 
did not.  We attribute 
the difference to the limitations of the $\mathcal{F}_{\rm xc}$
functional used in the latter work.  
$\Delta P_{\mathrm{tot}}$ is largest at the lowest densities, with a
range of about 6\% for both 0.20 and 0.506 g/cm$^3$.  That range
decreases to about 3\% (still with both signs) at $\rho_D = 1.9631$
g/cm$^3$, then it continues down to about 2\% at $\rho_D = 4.04819$
g/cm$^3$ and is 1\% at most for $\rho_D = 10.0$ g/cm$^3$.  (For
reference, the equilibrium simple cubic density at $T \approx 0$K is
about 1 g/cm$^3$.)  Of course, the relative pressure shift is
practically negligible at low $T$ because the low-$T$ limit of KSDT
was designed to recover the ground-state LDA. (Ref.\ \cite{LSDA-PIMC}
gives a comparison of the KSDT and PZ correlation energies at $T=0$
K.)

The overall behavior is clear.  $T$-dependent XC first raises the
pressure, then, as $T$ increases, it lowers the pressure compared to
that generated by a ground state XC before both go to a common
high-$T$ limit. That limit occurs at much higher $T$ than what is 
shown in Figs. \ref{P-T.H128R0.1}-\ref{P-T.D216R10.0}.
That limiting behavior occurs even though the
two approximate functionals, KSDT and PZ, have different 
high-$T$ limits.  But the XC contribution becomes negligible 
compared to the non-interacting free-energy contribution at high-$T$,
so the effect of those different limits is suppressed.   
There is some intermediate $T$ at which there is no shift between
the two functionals (see discussion of Fig.\ \ref{a-B-vs-T}
 below; also see Ref.\ \onlinecite{KarasievSjostromTrickey12B}). 
The well-defined maxima of the total pressure relative differences occur  
near $T\approx 40$ kK,
60 kK, 100 kK, 125 kK, and 200 kK for $r_{\mathrm{s}}=3$, 2.20, 1.40, 1.10, and 0.81373 bohr
respectively, with corresponding 
values of about $3$, $4$, $2.5$, $2$ and $1$ \%.
Note also the 
nice correlation of the XC thermal effect upon the pressure 
with Fig.\ \ref{fxc-map}.  The maximum effect occurs approximately 
along the lower edge of the yellow-orange band
and the maximum magnitude decreases with decreasing $r_{\mathrm{s}}$.

We note also that the OFDFT finite $T$ results at high $T$ are
in good overall agreement with PIMC simulation data 
\cite{Hu.Militzer..2011}. The PIMC calculations should describe the 
$T$-dependence of all free-energy terms correctly, including 
the electron-electron interaction and therefore, the XC free energy.
Figs. \ref{P-T.H128R0.1} - \ref{P-T.D216R10.0} demonstrate that inclusion 
of the $T$-dependent XC provides overall better agreement between the 
KS and PIMC data than does use of ground-state XC. The exception is  
points where PIMC clearly exhibits irregular behavior.  
That occurs at low-$T$ for some material densities, 
with the PIMC pressures  
seeming to be significant overestimates relative to  
the KS results for $r_{\mathrm{s}}=3$ bohr (see Fig. \ref{P-T.H128R0.1}),
$r_{\mathrm{s}}=2.20$ bohr (see Fig. \ref{P-T.D128R0.506}), 
and  $r_{\mathrm{s}}=1.40$ bohr (see Fig.\ \ref{P-T.D128R1.96361}).
For $r_{\mathrm{s}}=0.81373$ bohr the PIMC pressure is low relative to KS at 
the lowest available temperature $T=125$ kK.

\subsection{Hugoniot of liquid Hydrogen}
\label{hugoniot}

Experimentally the EOS at high compressions is accessible 
via shock-wave techniques and the Hugoniot relation
\be
{\mathpzc E}-{\mathpzc E}_0-{\half}(P+P_0)
\Big(\frac{1}{\rho}-\frac{1}{\rho_0}\Big)=0
\,,
\label{HugEq}
\ee
where ${\mathpzc E}$, $P$, and $\rho$ are the specific {\it internal} energy,
pressure, and bulk density of a state derived by shock compression from 
an initial state at $\rho_0$, ${\mathpzc E}_0$, and $P_0$.

The initial state presents some technical challenges for computation.   
To enable a meaningful comparison between energies of 
states calculated from different codes (and possibly with different 
pseudopotentials), ${\mathpzc E}_0$ and ${\mathpzc E}$
usually are calculated as effective atomization energies of the system.
Doing so provides some error cancellation, especially for approximate 
treatment of core electronic states.   
Additionally, zero-point vibrational energy (ZPE) corrections are 
needed.  For Hydrogen, the result is that the initial state specific energy  
takes the form
\be
{\mathpzc E}_0=
\frac{E({\mathrm{H}}_N)+NE_{\mathrm{vib}}/2-N E(\mathrm{H})}
{Nm_{\mathrm{H}}}
\,,
\label{E0}
\ee
where $E_{\mathrm{vib}}$ is the ZPE for the H$_2$ diatomic molecule,
$E({\mathrm{H}}_N)$ is the energy of the $N$-atom system corresponding
to the initial conditions at material density $\rho_0$ and temperature $T_0$,
 $E(\mathrm{H})$ is the energy of an isolated H atom of mas $m_H$.  Note that 
$E(\mathrm{H})$ can be from a spin-polarized or non-spin-polarized
calculation, because eventually these terms cancel in Eq.\ (\ref{HugEq}).
Table \ref{tab:table2} shows
atomization energies ($D_e=2\{E_{np}({\mathrm{H}}_N)-N E_{np}(\mathrm{H})\}/N$) from
the non-spin-polarized calculation ($E_{np}$),
and values of pressure and energy ${\mathpzc E}_0$ corresponding to the initial state with
$\rho_0=0.0855$ g/cm$^3$ at $T_0=20$ K (essentially equilibrium bulk
H$_2$).    
In terms of $D_e$, the specific energy of the initial state is 
given by ${\mathpzc E}_0=(D_e+E_{vib})/2m_{\mathrm{H}}$.  The vibrational
correction is from the theoretical ZPE
obtained from DFT vibrational frequency calculations for the H$_2$ molecule
with the aug-cc-pVQZ basis set \cite{Dunning.1989-1992}, using 
the Vosko-Wilk-Nusair LDA \cite{VWN80} and Perdew-Burke-Ernzerhof 
(PBE)\cite{PBE96} GGA functionals.  
We remark that 
${\mathpzc E}_0$ values in Table \ref{tab:table2} are shifted by $E_{np}(\mathrm{H})/m_{\mathrm{H}}$ 
with respect both to $E({\mathrm{H}}_{64})/64m_{\mathrm{H}}$ (see Eq.\ (\ref{E0}))
and to the value reported in Ref.\ \cite{Kolos.1964} and used in Ref.\ \cite{DanelKazandjianPiron16}.
For example, for 32 H$_2$ molecules at initial conditions used here, 
our QuantumEspresso calculation with PZ XC gives -31.009 eV/molecule
or -15.505 eV/atom. Correcting by the 0.260 eV/molecule ZPE
gives -15.375 eV/atom, equivalent to ${\mathpzc E}_0=-1472$ kJ/g 
unshifted value.

\begin{table}
\caption{\label{tab:table2}
Pressure (kBar), atomization energy $D_e$ (eV/molecule),
and corresponding ZPE-corrected ${\mathpzc E}_0$ (in kJ/g) 
obtained from MD simulations for Hydrogen at $\rho_{0}=0.0855$ g/cm$^3$, $T=20$ K 
with different codes/functionals. ``QE'' is Quantum Espresso. 
All cases used PAWs.  
}
\begin{ruledtabular}
\begin{tabular}{lcccc}
Code &   XC  & $P_0$ & $D_{e}$\footnotemark[1]     & 
                       ${\mathpzc E}_{0}$ 
\\
\hline
QE     & PZ   & -2.2 & -6.7370 &  -310.0\footnotemark[2] \\ 
QE     & KSDT & -2.3 & -6.7264 &  -309.5\footnotemark[2] \\ 
QE     & PBE  & 0.25 & -6.7703 & -311.3\footnotemark[3] \\
\hline     
VASP   & PBE  & 0.21 & -6.7756 & -311.5\footnotemark[3] \\
VASP   & PBE  & && -314\footnotemark[4] \\
\end{tabular}
\end{ruledtabular}
\footnotetext[1]{$D_{e}=(E_{np}(H_{64})-64E_{np}(H)$)/32.}
\footnotetext[2]{ZPE correction $E_{vib}=\omega_e^{\mathrm{LDA}}/2=0.260$ eV.}
\footnotetext[3]{ZPE correction  $E_{vib}=\omega_e^{\mathrm{PBE}}/2=0.267$ eV .}
\footnotetext[4]{Ref. \onlinecite{Holst08}.}
\end{table}

\begin{figure}
\includegraphics*[width=7.5cm]{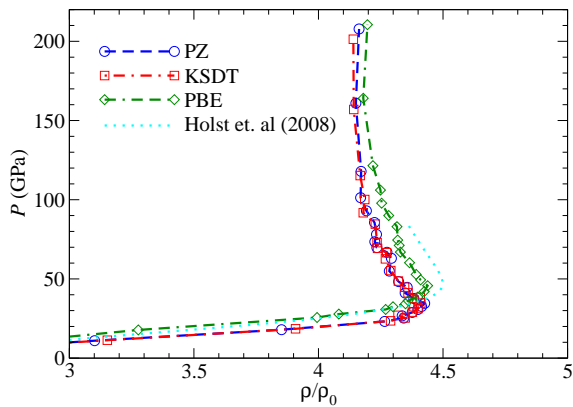}
\caption{
Hydrogen principal Hugoniot.   Initial
density $\rho_0=0.0855$ g/cm$^3$. 
}
\label{H-Hug}
\end{figure}

\begin{figure}
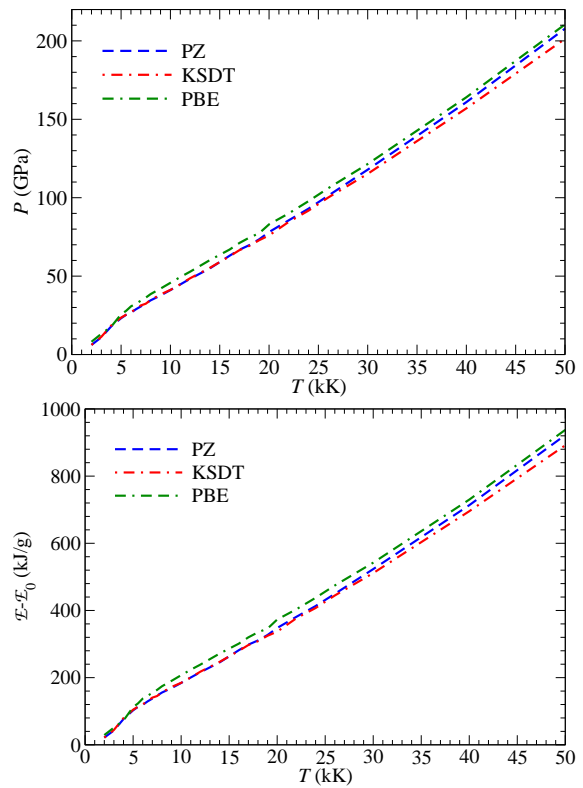

\includegraphics*[width=7.5cm]{Hug.P-vs-T.PZ-KSDT-PBE.ZPE0.260-0.267eV.v3b.eps}
\includegraphics*[width=7.5cm]{Hug.E-vs-T.PZ-KSDT-PBE.ZPE0.260-0.267eV.v3b.eps}
\caption{
Pressure (upper panel) and specific {\it internal} energy difference (lower panel)
along the Hydrogen Hugoniot as functions of $T$.
}
\label{H-Hug3}
\end{figure}

Figure \ref{H-Hug} compares the Hydrogen principal Hugoniot 
from the simulations with the KSDT and PZ XC functionals. Results
for the PBE  
GGA XC (also with $T$-dependent density) are shown to provide an 
alternative perspective on the effects of
changing only the XC approximation.  For $T < 30$ kK, there is
little XC thermal effect.   For example, the maximum compression is 4.41 for 
KSDT versus 4.43 for PZ at $P\approx 35$ GPa.  Shifting from LDA to GGA
(both ground-state functionals, PZ vs.\ PBE) changes the 
result only to 4.44 (PBE) but at notably higher pressure, $P\approx 46$ GPa.
For $T\ge 30$kK ($P\ge 120$ GPa), $P$ and $T$ increase practically 
at constant compression for all three curves. 
The $T$-dependent XC predicts slightly lower pressures than those from  
PZ,  in agreement with the results shown in Sec.\ \ref{EOS}.
This can be seen in the upper panel of Fig.\ \ref{H-Hug3}, 
which displays $P(T)$ along the Hugoniot. At $T=30$ kK the effects
on $P$ of $T$-dependence in XC versus shifting to 
gradient corrections in XC are comparable.
As $T$ increases, gradient corrections diminish in importance and the PZ 
and PBE curves become closer. In contrast, 
the effect of explicit $T$-dependence continues to increase.
The lower panel of Fig.\ \ref{H-Hug3} shows the same comparison for 
the specific {\it internal} energy (relative to the reference state). 
At low $T$, the KSDT internal energy is slightly higher than the PZ
result whereas at high $T$ the reverse is true.  Overall the
two yield nearly identical values.  That helps explain
why the Hugoniot curve, Fig.\ \ref{H-Hug}, is insensitive
to the use of KSDT rather than PZ XC. In the region of primary
interest, KSDT lowers both 
quantities on the LHS of Eq.\ \ref{HugEq}, 
$P$ and ${\mathpzc E}$, relative to PZ XC values, such that the solution, 
the material density $\rho$, 
remains almost unchanged  as compared to PZ XC results.
This insensitivity of the Hugoniot to $T$-dependence in XC
agrees with the findings of 
Ref.\ \onlinecite{Tubman..Ceperley.PRL2015}, namely that the Hugoniot is 
determined mainly 
by the statistics of nuclear configurations, not by the electronic $T$.

\subsection{Homogeneous and in-homogeneous electron gas at finite {\it T}}
\label{electron_gas}

Insight into the behaviors discussed in the preceding sections may
be gained by going back to basics, namely the HEG.  
The KSDT functional Eq.\ (\ref{fit2}) is itself an accurate 
parametrization of RPIMC simulation data for the finite-$T$ HEG. 
Closely related to the HEG is static bulk atomic H, a hypothetical 
system we have used to test OFDFT non-interacting free-energy functionals.  
The system is an abstraction of the   
experimental coexistence of hot electrons and cold ions that 
can occur with femtosecond laser pulses \cite{Ernstorfer.Science.2009}. \\[-5pt]

\begin{figure}
\includegraphics*[width=7.5cm]{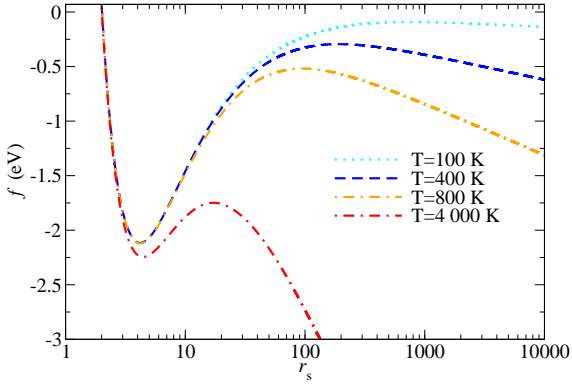}
\caption{
HEG total free-energy per electron as a function of $r_{\mathrm s}$
for selected temperatures calculated with the KSDT XC functional.
}
\label{ftot}
\end{figure}

\begin{figure}
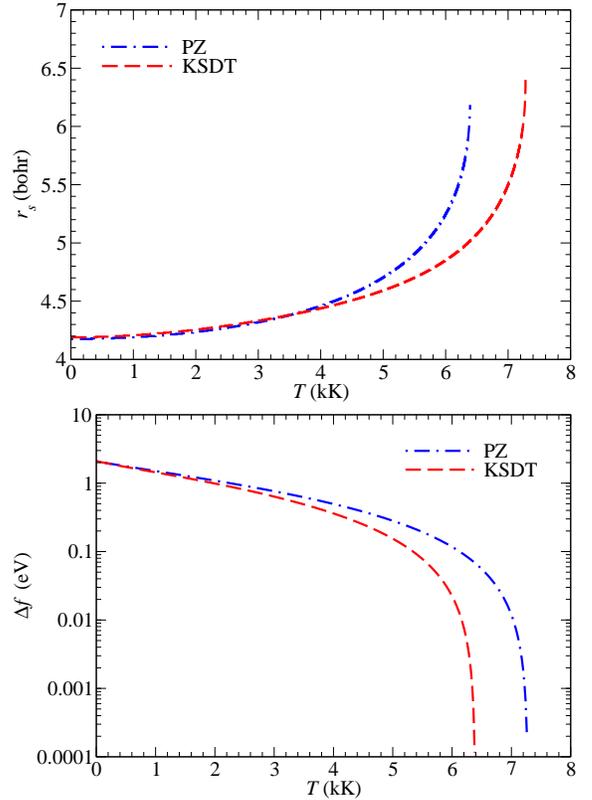

\includegraphics*[width=7.5cm]{rs_equil-vs-T.v2.eps}
\includegraphics*[width=7.5cm]{fbind-vs-T.v2.eps}
\caption{
Upper panel: value of equilibrium $r_{\mathrm{s}}$ corresponding to the
minimum of the total free-energy per electron for the HEG 
as a function of $T$.
Lower panel:  HEG barrier height (binding energy; see Fig. \ref{ftot} and
see text) as function of $T$. 
}
\label{rs_equil}
\end{figure}

For the HEG, consider first its bulk equilibrium density as a function
of $T$, i.e., that value of $r_{\mathrm s}$ 
for which the HEG free energy per particle is minimum.  
Figure \ref{ftot} shows this free energy per particle ($f=f_{\mathrm{s}}+f_{\mathrm{xc}}$)
as a function of $r_{\mathrm{s}}$ for selected temperatures calculated with the KSDT functional.
The upper panel of Fig.\ \ref{rs_equil} shows the difference for equilibrium $r_{\mathrm{s}}$
between KSDT and PZ. At $T=0$ K the equilibrium $r_{\mathrm{s},equilib}=4.19$ bohr
for both XC functionals (see also Ref.\ \onlinecite{Perdew.Wang.1992}).
The value 
remains almost constant up to $T\approx 1000$ K.
The ground-state PZ result starts to deviate from the finite-$T$ KSDT 
values at $T\approx 4000$ K.
The HEG explodes, in the sense that the $r_{\mathrm{s},equilib}$ 
increases drastically 
at about  $T_{\mathrm{c}}\approx 7200$K 
for the finite-$T$ XC. Use of the ground-state PZ XC approximation lowers
that substantially, to about 6200K.  
What may be construed as the HEG binding energy is shown in the 
lower panel of Fig.\  \ref{rs_equil}.  The quantity $\Delta f$ shown there
is the depth of the minimum of the total free-energy per particle (see Fig. \ref{ftot}) relative to the maximum at lower density (higher $r_{\mathrm s}$).  
As Fig.\ \ref{ftot} shows, one may also construe $\Delta f$ as a barrier
height.  
For both KSDT and PZ XC, $\Delta f$ decreases with increasing $T$
starting from 1 eV at $T$=0 K and plunging to essentially zero at the 
same critical values of $T$ as before, about 7200 K for KSDT vs.\ 6200 K for PZ,
a 14\% shift.  Given the structureless nature of the HEG, these comparisons 
drive home the point that the low density regime is 
rather sensitive to XC thermal effects.
 \\

\begin{figure}
\includegraphics*[width=7.5cm]{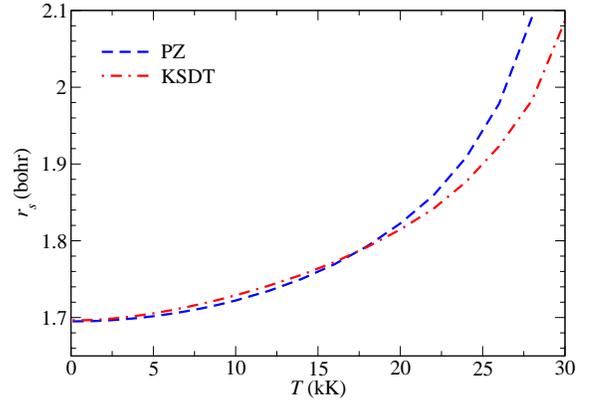}
\caption{
%
Equilibrium $r_{\mathrm{s}}$
as a function $T$ 
for sc-H. See text.  
}
\label{a-B-vs-T}
\end{figure}

Effects of reduction in translational invariance to periodic are 
illuminated by considering simple cubic bulk atomic Hydrogen 
(``sc-H'' hereafter).  
In essence, it is an inhomogeneous electron gas with the simplest
conceivable point charge compensating
background. 
Figure \ref{a-B-vs-T} shows the equilibrium $r_{\mathrm{s}}$ 
as a function of $T$. 
The behavior is similar to that for the HEG, namely a monotonic 
increase of the equilibrium  $r_{\mathrm{s}}$  with increasing 
$T$ and substantially lower values of the equilibrium $r_{\mathrm{s}}$ 
from the $T$-dependent KSDT XC than from PZ XC 
at high-$T$ (20 kK $<$ $T$ $<$ 30 kK). 
For 5 kK $<$ $T$ $<$ 15 kK the situation reverses, with KSDT 
giving slightly larger
$r_{\mathrm{s}}$ values  than PZ. Thus either the pressures 
or the equilibrium  $r_{\mathrm{s}}$ values from the two XC approximations
will coincide at some intermediate $T$.  Such behavior was observed 
previously  for  bcc-Li (see Fig.\ 11 in Ref.\ 
\onlinecite{KarasievSjostromTrickey12B}) and is consistent with
the AIMD results discussed above (recall Section \ref{EOS}).
For $T$ $>$ 30kK the sc-H model becomes unstable.
Replacement of the uniform background in the case of HEG with 
compensating point charges in sc-H makes the average equilibrium density 
at $T=0$ K of the sc-H much higher than for the 
HEG ($r_{\mathrm{s},equilib}=4.19$ bohr for the HEG vs.\ 1.70 bohr for sc-H ), 
hence sc-H is stable to much higher
$T$ ($T_{\mathrm{c},{\mathrm sc-H}}\approx$ 30 000 K vs.\   
$T_{\mathrm{c},{\mathrm HEG}}\approx$ 7200 K).  Compare
Figs.\ \ref{rs_equil} and \ref{a-B-vs-T}. 

\begin{figure}
\includegraphics*[width=7.5cm]{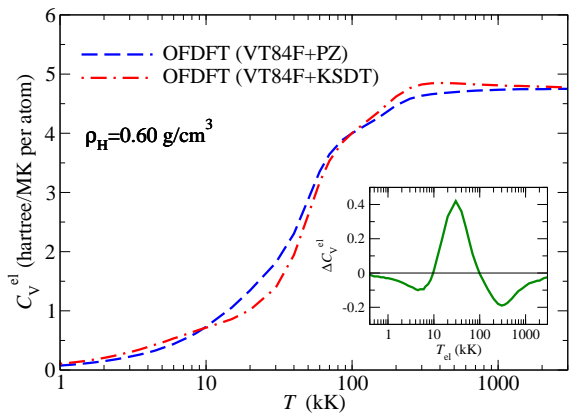}
\caption{
Electronic heat capacity at constant volume $C_V^{el}$ as a function
of electronic temperature for sc-H at material density 
$\rho_{\mathrm{H}}$ = 0.60 g/cm$^3$
with the finite-$T$ KSDT and ground-state PZ 
 XC functionals. The inset shows the difference
$\Delta C_V^{el}=C_V^{el,\mathrm{PZ}}-C_V^{el,\mathrm{KSDT}}$.
}
\label{C_v}
\end{figure}

Finally, we consider the sc-H electronic heat capacity at 
constant volume as a function of electronic temperature $T$. 
\be
C_V^{el}=(\partial {\mathcal E}^{el}/\partial T)_V  \; .
\label{Cv}
\ee
It obviously is a measure of the $T$-dependence of the 
electronic internal energy ${\mathcal E}^{el}$, which of course has an XC 
contribution ${\mathcal E}_{\mathrm{xc}}$.  
That $T$-dependence is qualitatively different for the zero-$T$ and finite-$T$
XC functionals (see Refs.\ \onlinecite{KarasievSjostromTrickey12B,LSDA-PIMC}).
$C_V^{el}$ therefore provides a different direct measure of XC thermal
 effects from that provided by quantities considered thus far.    
Figure \ref{C_v} compares $C_V^{el}$ 
obtained from the KSDT and PZ XC functionals for sc-H. These were done with
OFDFT using the VT84F non-interacting free-energy functional \cite{VT84F}. 
The maximum magnitude of the difference 
$\Delta C_V^{el}=C_V^{el,\mathrm{PZ}}-C_V^{el,\mathrm{KSDT}}$
for $\rho_{\mathrm{H}}$ = 0.60 g/cm$^3$
is 0.4 hartree/MK at $T\approx 30$ kK. That corresponds to about a
40\% relative difference.  In the zero-$T$ and high-$T$ limits the
difference between the KSDT and PZ XC results vanishes as expected.
Notice how those limits lead to broad consistency of behavior with
respect to other quantities discussed above.  XC thermal effects
relative to a ground-state functional are quite variable in magnitude.
For $C_V^{el}$ the difference is manifested both as a maximum and as
two other extrema of lesser magnitude.  \\

\section{Concluding remarks}
\label{conclusions}

The increase in resistivity of Al at 0.1 g/cm$^3$ found in
Ref.\ \onlinecite{Renaudin..Noiret.2002}
upon going from the VWN ground-state XC functional
\cite{VWN80} to the Perrot-Dharma-wardana (PDW) $T$-dependent functional
\cite{PDW84} is qualitatively consistent with our finding
of the lowered DC conductivity of low-density Al.  Quantitatively,
however, the results are quite different.  In particular, the low-$T$
behavior of the resistivity calculated from the PDW functional is
drastically different from our KSDT result, their VWN result, and 
Perrot and Dharma-wardana's earlier calculation \cite{PDW1987}.  
We suspect methodological differences but can not say more.

Ref.\ \onlinecite{OurFrontiersInWDM.2014} included an analysis
suggesting that intrinsic $T$-dependence in ${\mathcal F}_{\mathrm{xc}}$ did
not necessarily mean there would be a corresponding impact on the
$T$-dependence of the Kubo-Greenwood optical conductivity.  The nub of
that argument was that the conductivity depends on the KS eigenvalues
and orbitals, which are determined by the KS potential (for which the
XC contribution is $v_{\mathrm{xc}}= \delta {\mathcal F}_{\mathrm{xc}}/\delta n$). Sums
of matrix elements over such KS quantities do not necessarily have a
strong sensitivity to $T$-dependence of ${\mathcal F}_{\mathrm{xc}}$.  

Such arguments tend to overlook state conditions.  Here, we have given
an example of a system for which $\sigma_{\mathrm{DC}}$ has a stronger
sensitivity (in the sense of percentage shift) to XC thermal effects
in a pertinent range of thermodynamic conditions than does the EOS of
a different system over a different but also pertinent range of
thermodynamic conditions. This shows that both system definition and
state conditions are essential for proper assessment of the GSA.

For the DC conductivity of low-density Al, XC thermal effects
incoporated in the KSDT functional increase inter-band separations
relative to GSA above the Fermi level, hence decrease the Fermi-Dirac
occupations relative to GSA.  That lowers the calculated conductivity
(see Eq.\ (\ref{KGformulae})), and thereby yields better agreement
with the experimental data. We remark that the ground-state
functionals underestimate band gaps, while hybrid functionals (which
incorporate a fraction of exact single-determinant exchange, thus take
into account a part of exchange thermal effects) enlarge such
intervals.  Therefore one expects that a hybrid functional would lower
the conductivity.  Notice that in the case of thermal Hartree-Fock,
thermal occupation at temperatures of 1 or 2 eV {\it reduces} the
spuriously large inter-band separations of ground-state HF states
\cite{KarasievSjostromTrickey12B}.  That, in turn, would drive  
the opposite trend, {\it increasing} the conductivity.  Interestingly, 
the overall XC thermal effect upon the DC conductivity of low density Al is to
reduce the range of the $T$-variation at fixed bulk density.
In contrast, at 
higher material density the Al DC conductivity is essentially insensitive 
to the XC $T$-dependence. 

The Deuterium EOS calculations show that XC thermal effects 
must be taken into account in thermodynamic 
conditions corresponding approximately to the 
reduced temperature $t\approx 0.5$.  
However, because the principal Hugoniot characterizes the  
difference of two states separated by a shock, there is a 
cancellation that is familiar in other uses of KS-DFT (e.g.,
atomization energies) which substantially suppresses the XC 
thermal effects.  As noted above, this cancellation
is consistent with the findings of Tubman {\it et al.}\  
\cite{Tubman..Ceperley.PRL2015} and with Danel {\it et al.} 
\cite{DanelKazandjianPiron16}.  We suspect therefore  
that the XC thermal 
effects on the Al Hugoniot found in Ref.\ \onlinecite{Surh2001} 
and on the Be Hugoniot in Ref.\ 
\onlinecite{Sterne..Isaacs.2007} are consequences of the techniques they 
used.  In the former work $T$-dependence was introduced by 
adding jellium shifts to the energy and pressure 
at $r_{\mathrm s}$ corresponding to the
density at hand.  The second used an average atom. 
Nevertheless the XC thermal effects on
the pressure $P(T)$ (recall upper panel of Fig.\ \ref{H-Hug3})
are not negligible.  Rather the pressure effect is about
the same magnitude as the shift from gradient-independent to
gradient-dependent XC. This also is consistent with 
what is reported by Danel {\it et al.} 
\cite{DanelKazandjianPiron16}.  More accurate predictions 
(for all properties affected by the XC thermal effects)
will require both explicit $T$-dependence and gradient contributions in
the XC functional, at least.

Even at the LDA level of refinement, however, it is
clear that the GSA (use of a ground-state XC functional as an approximate 
free-energy XC functional) is not an unequivocally valid prescription
 \cite{Burke.Entropy}.   That assessment is consistent with
earlier demonstrations of the non-trivial $T$-dependence 
of the XC free-energy 
\cite{SjostromHarrisTrickey12,KarasievSjostromTrickey12B,LSDA-PIMC}.  
It also confirms what one knows in principle, namely that consistent
study of WDM requires an approximate ${\mathcal F}_{\mathrm{xc}}$ which has
a proper  high-$T$ limit, a
correct small-$r_{\mathrm{s}}$ limit, a correct 
small-$\Gamma$ (the dimensionless Coulomb coupling parameter) limit,
and delivers a properly positive entropy.  The KSDT functional is
built with the first three included explicitly and is found to 
satisfy the last {\it a posteriori} in every case considered. 

\section{Acknowledgments}
We thank Debajit Chakraborty and Travis Sjostrom for helpful comments
on the manuscript and Jim Dufty and Mike 
Desjarlais for helpful discussions regarding optical 
conductivity calculations.  We thank Kieron Burke for providing the
revised version of Ref.\ \onlinecite{Burke.Entropy} and for alerting
us to the entropy negativity discussed in Sect.\ \ref{entropy-constraint}.  We thank an
anonymous referee for noting the expected effect of hybrid XC 
functionals upon interband separations and, thereby, the DC conductivity.  
Our work was supported by U.S.\ Dept.\ of Energy grant DE-SC0002139. 
We thank the University of Florida Research Computing organization 
for computational resources and technical support. 
The most demanding optical conductivity calculations were feasible 
only because of resources provided by the National 
Energy Research Scientific Computing Center, 
a DOE Office of Science User Facility  
supported by the Office of Science of the U.S.\ Department of Energy 
under Contract No. DE-SC0002139.


\end{document}